\documentclass[12pt]{article}

\usepackage{amsmath,amsfonts,amssymb}

\def\bA{\mathbf{A}}

\def\balpha{\bar{\alpha}}
\def\bbeta{\bar{\beta}}

\def\be{\begin{equation}}

\def\ee{\end{equation}}

\def\bea{\begin{eqnarray}}

\def\eea{\end{eqnarray}}

\def\bgamma{\bar{\gamma}}
\def\bdelta{\bar{\delta}}

\def\mH{\mathcal{H}}

\newcommand{\mF}{\mathcal{F}}

\newcommand{\mG}{\mathcal{G}}

\def \bA{\mathbf{A}}

\newcommand{\mL}{\mathcal{L}}

\begin{document}

\begin{titlepage}

\vskip 0.4 cm

\begin{center}
{\Large{ \bf Canonical Analysis of Geometrical Action
for  Dp-brane}}

\vspace{1em}  Josef Kluso\v{n}$\,^1$
\footnote{Email address:
 klu@physics.muni.cz}\\
\vspace{1em} $^1$\textit{Department of Theoretical Physics and
Astrophysics, Faculty
of Science,\\
Masaryk University, Kotl\'a\v{r}sk\'a 2, 611 37, Brno, Czech Republic}\\

%
%

\vskip 0.8cm

\end{center}

\begin{abstract}
In this short note we perform canonical analysis of geometrical action for
 Dp-brane. We also discuss tachyon condensation in case of the geometrical
action for unstable D(p+1)-brane.

\end{abstract}

\bigskip

\end{titlepage}

\newpage

\section{Introduction and Summary}
It is well known that effective action for Dp-brane is
Dirac-Born-Infeld (DBI) action
\cite{Polchinski:1995mt,Leigh:1989jq}(For review, see for example
\cite{Simon:2011rw}). The problem with this action is that it is
non-polynomial in world-volume coordinates and in world-volume gauge
field that makes its analysis rather complicated. Similar situation
occurs with fundamental string whose dynamics is governed by
Nambu-Gotto form of the action which is again non-polynomial in
string coordinates. On the other hand it is well known that it is
possible to replace this action with classically equivalent form of
the action with world-sheet metric
\cite{Brink:1976sc,Deser:1976rb,Howe:1977hp}. An analogue procedure
can be performed in case of $p-$brane
\cite{Howe:1977hp,AbouZeid:1997mt}. On the other hand situation is
more involved in case of DBI action due to the presence of the gauge
field so that auxiliary metric contains symmetric and anti-symmetric
part \cite{AbouZeid:1997mt,Lindstrom:1987cv}. On the other hand an
interesting form of the action for Dp-brane was proposed in
\cite{AbouZeid:1998he} which is known as geometrical form of the
action. The main advantage of this action is that it depends on
quadratic combination of field strength and hence we can introduce
auxiliary metric which is symmetric.

Since the form of the action is very interesting and not well known
we mean that it deserves to be studied further. In this note we
focus on the canonical analysis of the geometrical action as was
introduced in \cite{AbouZeid:1998he}. We would like to see whether
Hamiltonian is different from the Hamiltonian of DBI action which is
given as linear combination of $p+1$ first class constraints. In
case of geometrical action the canonical analysis is more
complicated due to the form of Lagrangian density but we again find
that this theory has $p+1$ first class constraints which has the
same form as constraints derived from DBI action. This is certainly
nice consistency check that demonstrates that different Lagrangian
formulations that are in some way equivalent have the same
Hamiltonian. Well known example is Nambu-Gotto and Polyakov form of
the string action.

We also briefly discuss geometrical form of unstable D(p+1)-brane action
\cite{Sen:1999md,Bergshoeff:2000dq,Kluson:2000iy} when we easily generalize
analysis that leads to geometrical action. We also study the tachyon condensation in the form of the tachyon kink which is profile of the tachyon field that depends on one
world-volume coordinate. We study this problem following very nice analysis presented in
\cite{Erkal:2009xq} where it was argued that it is possible to interpret non-BPS D(p+1)-brane as $p+1$-dimensional object moving in $11$ dimensional space-time where additional coordinate corresponds to the tachyon field $T$. In fact, such a structure is also manifest in the geometrical form of unstable D(p+1)-brane action studied here. Then the tachyon kink corresponds to the partial gauge fixing when one world-volume coordinate can be identified with the tachyon field. Generally all world-volume fields on this gauge fixed form of the action depend on the tachyon. Then when we restrict ourselves to the low energy effective action  we can presume that all world-volume fields do not depend on tachyon we find that resulting action corresponds to the geometrical form of Dp-brane action. Situation when
the fluctuations modes depend on $T$ was nicely analyzed in \cite{Erkal:2009xq} and it was argued there that they are not normalizable and hence cannot be considered as open strings excitations. Rather they can be interpreted as excited closed string states and we recommend the original paper \cite{Erkal:2009xq} for more details.

Let us outline our result. We found canonical structure of the geometrical action and we showed that it has the same form as in case of the Hamiltonian analysis of DBI action. We also found geometrical form of unstable D(p+1)-brane action and studied tachyon condensation on its world-volume where we showed that it leads to the geometrical form of stable Dp-brane which is nice consistency check of the tachyon condensation.

\section{Canonical Formalism of Geometrical Action for Dp-brane}
To begin with we review construction of geometrical action that
was performed in \cite{AbouZeid:1998he}. We start with the standard DBI action
for Dp-brane that has the form
\begin{equation}
S=-T_p \int d^{p+1}\xi e^{-\phi}\sqrt{-\det (g_{\alpha\beta}+\mF_{\alpha\beta})} \ ,
\end{equation}
where $T_p=\frac{1}{l_s^{p+1}}$ is Dp-brane tension where $l_s$ is string length. Further,
$g_{\alpha\beta}=G_{MN}\partial_\alpha x^M\partial_\beta x^N$ where $G_{MN}$ is background metric, $x^M,M,N=0,1,\dots,9$ parameterize embedding of Dp-brane in the target space-time. Further, world-volume of Dp-brane is parameterized by coordinates $\xi^\alpha,\alpha,\beta=0,1,\dots,p$, where $\partial_\alpha=\frac{\partial}{\partial \xi^\alpha}$. $\phi$ is the background field known as dilaton and $\mF_{\alpha\beta}=b_{\alpha\beta}+l_s^2F_{\alpha\beta}
\ , F_{\alpha\beta}=
\partial_\alpha A_\beta-\partial_\beta A_\alpha$, where $A_\alpha$ is the gauge field that propagates on the world-volume of Dp-brane. Finally $b_{\alpha\beta}=B_{MN}\partial_\alpha x^M\partial_\beta x^N$ is an embedding of background NSNS two form to the world-volume of Dp-brane.

Geometrical form of the action was derived in \cite{AbouZeid:1998he} using of an important
fact that
\begin{equation}
\det (g_{\alpha\beta}+\mF_{\alpha\beta})=\det (g_{\alpha\beta}-\mF_{\alpha\beta})
\end{equation}
so that we can write
\begin{eqnarray}\label{defgeom}
& &\sqrt{-\det (g_{\alpha\beta}+\mF_{\alpha\beta})}=
(-\det (g_{\alpha\beta}+\mF_{\alpha\beta}))^{1/4}(-\det (g_{\alpha\beta}-\mF_{\alpha\beta}))^{1/4}=
\nonumber \\
& &=
(- g)^{1/4}(-\mG)^{1/4} \ , \quad g=\det g_{\alpha\beta} \ , \quad
\mG=\det \mG_{\alpha\beta} \ ,  \nonumber \\
& &\mG_{\alpha\beta}=
g_{\alpha\beta}-\mF_{\alpha\gamma}g^{\gamma\delta}\mF_{\delta\beta}
\ , \quad \mG_{\alpha\beta
}=\mG_{\beta\alpha} \ .
\nonumber \\
\end{eqnarray}
As a result we obtain geometrical form of the action
\begin{equation}\label{geoform}
S=-T_p \int d^{p+1}\xi e^{-\phi}(-g)^{1/4}(- \mG)^{1/4} \ .
\end{equation}
It was shown in   \cite{AbouZeid:1998he} that the  main advantage of
this action is that when we introduce  auxiliary world-sheet metric
we obtain an action that is quadratic in gauge fields. Further,
since the action (\ref{geoform}) is apparently different from DBI
form of the action it is interesting to study it in more detail. In
fact, in this note   we focus on the canonical analysis of the
action (\ref{geoform}).

Now from  (\ref{geoform}) we obtain conjugate momenta
\begin{eqnarray}\label{momen}
& &p_M=\frac{\delta \mL}{\delta \partial_0 x^M}=\nonumber \\
&&=
-\frac{1}{2}T_p e^{-\phi}g_{MN}\partial_\beta x^N
g^{\beta 0}(-g)^{1/4} (-\mG)^{1/4}
-\frac{1}{2}T_p g_{MN}\partial_\beta x^N \mG^{\beta 0}(-g)^{1/4}
(-\mG)^{1/4} -\nonumber \\
& &-\frac{1}{2}T_p g_{MN}\partial_\beta x^N g^{\beta\sigma}
\mF_{\sigma\omega}\mG^{\omega \delta}\mF_{\delta \rho}g^{\rho 0}
(-\mG)^{1/4}(-g)^{1/4}+\nonumber \\
& &+\frac{1}{2}T_p e^{-\phi}
(-g)^{1/4}(-\mG)^{1/4}(B_{MN}\partial_\beta x^N g^{\beta\rho}
\mF_{\rho\sigma}\mG^{\sigma 0}-B_{MN}
\partial_\beta x^N g^{0\rho}\mF_{\rho\sigma}\mG^{\sigma\beta}) \ , \nonumber \\
& &\pi^\alpha=\frac{\delta \mL}{\delta \partial_0 A_\alpha}=
\frac{l_s^2}{2}T_p e^{-\phi}(g^{\alpha\beta}\mF_{\beta\rho}\mG^{\rho 0}-
g^{0\beta}\mF_{\beta\rho}\mG^{\rho\alpha})(-g)^{1/4}(-\mG)^{1/4} \nonumber \\
\end{eqnarray}
so that $\pi^0\approx 0$. Further, the bare Hamiltonian density is equal to
\begin{eqnarray}
\mH=p_M\partial_0 x^M+\pi^iF_{0i}+\pi^i\partial_i A_0-\mL=\pi^i\partial_i A_0 \ .
\nonumber \\
\end{eqnarray}
To proceed further we use definition of $\pi^{\alpha}$ given in (\ref{momen}) to introduce
$\Pi_M$ defined as
\begin{eqnarray}
& &\Pi_M=p_M-l_s^{-2}B_{MN}\partial_i x^N \pi^i
=\nonumber \\
& &=
-\frac{1}{2}T_p e^{-\phi}g_{MN}\partial_\beta x^N
g^{\beta 0}(-g)^{1/4} (-\mG)^{1/4}
-\frac{1}{2}T_p g_{MN}\partial_\beta x^N \mG^{\beta 0}(-g)^{1/4}
(-\mG)^{1/4} -\nonumber \\
& &-\frac{1}{2}T_p g_{MN}\partial_\beta x^N g^{\beta\sigma}
\mF_{\sigma\omega}\mG^{\omega \delta}\mF_{\delta \rho}g^{\rho 0}
(-\mG)^{1/4}(-g)^{1/4} \ . \nonumber \\
\end{eqnarray}
Now using (\ref{momen}) or its equivalent form given above we get
\begin{eqnarray}
& &\partial_i x^M \Pi_M+F_{ij}\pi^j=
\partial_i x^M p_M+F_{ij}\pi^j=
\nonumber \\
& &=-\frac{1}{2}T_p (-g)^{1/4}g_{i\beta}\mG^{\beta 0}(-\mG)^{1/4}
-\frac{1}{2}T_p (-g)^{1/4}(-\mG)^{1/4}F_{i\omega}\mG^{\omega\delta}F_{\delta\rho}g^{\rho 0}
+\nonumber \\
& &+\frac{1}{2}T_p e^{-\phi}F_{ij}g^{j\beta}F_{\beta\rho}\mG^{\rho 0}(-g)^{1/4}(-\mG)^{1/4}
-\frac{1}{2}T_p e^{-\phi}F_{ij}g^{0\beta}F_{\beta\rho}\mG^{\rho j}(-g)^{1/4}(-\mG)^{1/4}=
\nonumber \\
& &=-\frac{1}{2}T_p e^{-\phi}(-g)^{1/4}(-\mG)^{1/4}
(g_{i\beta}-F_{i\gamma}g^{\gamma\rho}F_{\rho\beta})\mG^{\beta 0}=0 \nonumber \\
\end{eqnarray}
that implies an existence of $p-$primary constraints $\mH_i$ defined as
\begin{equation}
\mH_i=p_M\partial_i x^M+F_{ij}\pi^j\approx 0 \
\end{equation}
that are standard spatial diffeomorphism constraints.

As the next step we should find Hamiltonian constraint. To do this we have to use
crucial properties of the matrix $\mG$ as follows from its definition given in
(\ref{defgeom}). Namely, it is easy to see that
\begin{equation}
\mF_{\mu\nu}g^{\nu\rho}\mG_{\rho\sigma}=\mG_{\mu\nu}g^{\nu\sigma}\mF_{\sigma\rho}
\end{equation}
or in matrix notation
\begin{equation}\label{mFgmG}
\mF g^{-1}\mG=\mG g^{-1}\mF \ .
\end{equation}
Using this relation we obtain
\begin{eqnarray}
g\mG^{-1} \mF g^{-1}\mF=\mF\mG^{-1}\mF
\nonumber \\
\end{eqnarray}
that, in the end gives an important relation
\begin{eqnarray}\label{mG1}
\mG^{-1}-g^{-1}=g^{-1}\mF\mG^{-1}\mF g^{-1} \ , \nonumber \\
\end{eqnarray}
where we also used the fact that $\mF g^{-1}\mF=g-\mG$. Further, from (\ref{mFgmG}) we
obtain
\begin{eqnarray}\label{mFgmG1}
\mG^{-1} \mF g^{-1}=g^{-1}\mF \mG^{-1} \ . \nonumber \\
\end{eqnarray}
On the other hand using definition of $\mG$ we get
\begin{equation}
g^{\mu\nu}\mF_{\nu\rho}\mG^{\rho\sigma}=
-\mG^{\sigma\rho}\mF_{\rho\nu}g^{\nu\mu} \ .
\end{equation}
Now if we combine this relation with (\ref{mFgmG1}) we obtain
\begin{equation}\label{relanti}
\mG^{\mu\nu}\mF_{\nu\rho}
g^{\rho\sigma}=-\mG^{\sigma\rho}\mF_{\rho\nu}g^{\nu\mu} \ .
\end{equation}
Then with the help of these results we can simplify expressions for canonical momenta given
in (\ref{momen}) as
\begin{eqnarray}\label{PiM}
\Pi_M=-T_p e^{-\phi}g_{MN}\partial_\beta x^N\mG^{\beta 0}(-g)^{1/4}(-\mG)^{1/4} \ , \quad
\pi^\alpha=T_p e^{-\phi}g^{\alpha\beta}\mF_{\beta\delta}\mG^{\delta 0}(-g)^{1/4}(-\mG)^{1/4} \ .
\nonumber \\
\end{eqnarray}
Now we can proceed to the search for Hamiltonian constraint. We can expect that it
will be quadratic in momenta and so that it is natural to consider following combination
 $\Pi_M G^{MN}\Pi_N+\pi^ig_{ij}\pi^j$. Then, using (\ref{PiM}) we obtain
\begin{eqnarray}\label{searchcon}
\Pi_M G^{MN}\Pi_N+\pi^i g_{ij}\pi^j=-T_p^2 e^{-2\phi}\mG^{00}(-g)^{1/2}(-\mG)^{1/2} \ .
\nonumber \\
\end{eqnarray}
To proceed further let us again return to the definition of $\mG$ given in
(\ref{defgeom})
and write it in the form
\begin{eqnarray}
\mG_{\alpha\beta}=(g_{\alpha\gamma}+\mF_{\alpha\gamma})(\delta^\gamma_\beta-g^{\gamma\delta}\mF_{\delta\beta})
\nonumber \\
\end{eqnarray}
or in matrix notation
\begin{equation}
 \mG=(g+\mF)(I-g^{-1}\mF) \ .
 \end{equation}
Taking inverse of this relation and performing further manipulation we get
\begin{equation}
(I-g^{-1}\mF)\mG^{-1}=(g+\mF)^{-1}
\end{equation}
that implies following  relation
\begin{equation}
\mG^{-1}-g^{-1}\mF\mG^{-1}=(g+\mF)^{-1} \ .
\end{equation}
Now since $\mG^{0\alpha}\mF_{\alpha\beta}g^{\beta 0}=0$ as follows from (\ref{relanti})
for $\mu=\sigma=0$
we obtain important result
\begin{equation}
\mG^{00}=(g+\mF)^{00}=\frac{\det (g_{ij}+\mF_{ij})}{\det (g+\mF)} \ .
\end{equation}
Inserting this result into (\ref{searchcon}) we obtain final form of the
Hamiltonian constraint
\begin{equation}
\mH_\tau=\Pi_M G^{MN}\Pi_N+\pi^ig_{ij}\pi^j+T_p^2 e^{-2\phi}\det (g_{ij}+F_{ij})\approx 0 \ .
\end{equation}
We see that the Hamiltonian constraint has the same form as in case of DBI action. In summary, we find that the Hamiltonian formulation of the geometrical action for Dp-brane consists $p+1$ primary constraints $\mH_i,\mH_\tau$ that are first class constraints which simply follow from the fact that they have the same form as constraints that follow from  DBI action. Further, the requirement of the preservation of the primary constraint $\pi^0\approx 0$ implies secondary constraint $G\equiv \partial_i \pi^i\approx 0$ again with agreement with standard DBI action. In other words despite apparently different Lagrangian structure between geometrical and DBI actions we see that their Hamiltonian formulations  are the same.

\section{Unstable D(p+1)-brane}
The generalization of this approach to the case of unstable D(p+1)-brane is straightforward. To begin with we start with tachyon effective action
\cite{Sen:1999md,Bergshoeff:2000dq,Kluson:2000iy}
\begin{equation}
S=-\tau_{p+1}\int d^{p+2}\xi e^{-\phi}V(T)\sqrt{-\det \bA} \ ,
\end{equation}
where  $\bA_{\alpha\beta}=g_{\alpha\beta}+l_s^2\partial_\alpha T\partial_\beta T+
l_s^2\mF_{\alpha\beta}$ where $T$ is the tachyon field, $V(T)$ is tachyon potential with two minima $T_{min}=\pm \infty$ where $V(T_{min})=0$ and one local maximum $T_{max}=0$ where $V(T_{max})=1$
\footnote{For simplicity we restrict ourselves to the case of zero NSNS two form.}. Finally, $\tau_{p+1}$ is tension of unstable D(p+1)-bane.

In order to demonstrate an analogy between tachyon and additional target space coordinate
let us introducing variables $Y^I=(x^M,T)$ and
generalized metric $H_{IJ}, I,J=0,\dots,10$ in the form
\begin{equation}
H_{IJ}=\left(\begin{array}{cc}
G_{MN} & 0 \\
0 & l_s^2 \\ \end{array}\right)
\end{equation}
so that $h_{\alpha\beta}=\partial_\alpha Y^IH_{IJ}\partial_\beta Y^J=
\partial_\alpha x^M g_{MN}\partial_\beta x^N+l_s^2\partial_\alpha T\partial_\beta T$. Then it is easy to see that the geometrical action for non-BPS D(p+1)-brane has the form
\begin{equation}\label{unstablegeo}
S=-\tau_{p+1}\int d^{p+2}\xi e^{-\phi}V(-h)^{1/4}(-\mH)^{1/4} \ ,
\quad \mH_{\alpha\beta}=h_{\alpha\beta}-l_s^4 F_{\alpha\gamma}h^{\gamma\delta}F_{\delta\beta}
 \ .
\end{equation}
It is clear that the Hamiltonian analysis of this D(p+1)-brane is the same as in case of stable Dp-brane so that we will not repeat it here. On the other hand we would like to see that the tachyon kink solution corresponds to stable Dp-brane. We study this problem following \cite{Erkal:2009xq}. Explicitly, tachyon kink solution corresponds to the tachyon profile $T=f(z)$ where $z=\xi^{p+1}$ and where $f(z)$ is a function with $\frac{df}{dz}>0$ for all $z$. The simplest possibility is $f(z)=z$ and hence tachyon kink solution corresponds to the
gauge fixing in the extended space-time with the metric $H_{IJ}$. Clearly generally all world-volume fields still depend on $T$ through the inverse relation $z=f^{-1}(T)$. Further, we can take $A_z=0$ by $T-$dependent gauge transformations. Following \cite{Erkal:2009xq} and also \cite{Sen:2003tm} we consider situation when all world-volume fields
do not depend on $T$
\footnote{For general analysis, see \cite{Erkal:2009xq,Sen:2003tm}. Roughly speaking, it was argued in \cite{Erkal:2009xq} that $T-$dependent fluctuations are non-normalizable and hence cannot correspond to open string excitations. Rather they should be interpreted as creating of non-trivial closed string.}.
 Let us denote remaining world-volume variables as $\xi^{\balpha} , \balpha=0,1,\dots,p$ so that the matrix $h_{\alpha\beta}$ has the form
\begin{equation}\label{mhkink}
h_{zz}=l_s^2f'^2(z)  \ , \quad h_{\balpha\bbeta}=g_{\balpha\bbeta} \ , \quad h_{\balpha z}=0 \ .
\end{equation}
Further, the matrix $g^{\alpha\beta}$ is equal to
\begin{equation}
g^{\alpha\beta}=\left(\begin{array}{cc}
g^{\balpha\bbeta} & 0 \\
0 & \frac{1}{f'^2} \\ \end{array}\right)
\end{equation}
so that we obtain
\begin{eqnarray}\label{mHkink}
& &\mH_{zz}=l_s^2f'^2 \ , \quad  h_{zz}=f'^2 \ , \quad  h_{z\balpha}=0 \ ,  \nonumber \\
& & \mH_{z\balpha}=0 \ , \quad  h_{\balpha\bbeta}=g_{\balpha\bbeta} \ , \quad
\mH_{\balpha\bbeta}=g_{\balpha\bbeta}-l_s^4 F_{\balpha\bgamma}g^{\bgamma\bdelta}
F_{\bdelta \bbeta}\equiv \mG_{\alpha\beta} \ .  \nonumber \\
\end{eqnarray}
Inserting (\ref{mhkink}) and (\ref{mHkink}) into (\ref{unstablegeo})  we get
\begin{equation}
S_{non}^{fixed}(T=f(z))=-\tau_{p+1}l_s\int dz V(f(z))f'(z)
\int d^{p+1}\xi e^{-\phi}(-g)^{1/4}(-\mG)^{1/4}  \
\end{equation}
so that when we identify
\begin{equation}
T_p=\tau_{p+1}^{non}l_s\int dm V(m)
\end{equation}
we obtain an geometrical form of action for stable Dp-brane which is again nice consistency check of the tachyon condensation.

\end{document}